# Figuring and Drawing

## A Visual Approach to Principled Programming

Elpida Keravnou-Papailiou[a]

a Department of Computer Science, University of Cyprus, Cyprus

**Abstract** A standing challenge in undergraduate Computer Science curricula is the teaching and learning of computer programming. Through this paper which is an essay about programming, we aim to contribute to the plethora of existing pedagogies, approaches and philosophies, by discussing a specific feature of our approach in teaching principled programming to undergraduate students, in their first semester of studies, namely the utilization of pictures, both text-based and raster-based graphics. Although the given course has evolved substantially over the thirty years of its delivery regarding the programming languages (Miranda, C, C++, Java) and paradigms (functional, imperative, object-oriented, combination of procedural and object-oriented) used, the discussed visual feature has been maintained and steadily strengthened.

We list abstraction, problem decomposition and synthesis, information hiding, reusability, modularity and extensibility as key principles of problem solving and algorithmic thinking. These principles are closely aligned with the advocated computational thinking techniques of problem decomposition, pattern recognition, pattern generalization and algorithm design. We aim for our students to familiarize themselves with all the above principles through practical problem solving. Our ongoing inquiry has been whether the problem domain of pictures is contributing valuably towards this aim. Moreover, an added-value is that students get a glimpse of computational complexity in a visual, empirical way.

The presented work is not related to visual programming, since the students write their programs textually and not graphically; it's the output of their programs which is in visual form. Our approach though is loosely related to the classical paradigm of turtle graphics. However, our focus is Computer Science majors, who should be able to design and build turtles and other objects and not just use them. Indeed, the programming principles course helps them to do both and also to appreciate the multitude of algorithmic ways for producing the same visual output. Currently the given programming principles are approached both from a procedural, process-based and an object-oriented, concept-based perspective and the course uses the Java language. Through the presented example problems, we aim to show the appropriateness of the visual domain of pictures for supporting the learning of principled programming. The problem domain of pictures is abundantly rich with potential examples to draw from. Moreover, as reported in the literature, female students may show higher interest towards visual problem domains in programming classes, in relation to other problem domains. We plan to investigate this conjecture in the context of our broader aim to encourage more females to follow university studies in computer science; in this paper only a cursory finding is presented, that bears some relation to what is reported in the literature.

## 1 Introduction

This paper is an essay about programming, based on our thirty-year experience in teaching principles of programming to undergraduate students in Computer Science at the University of Cyprus, during their first semester of studies. Although the programming languages and associated paradigms have changed, a feature of the given course that has been maintained and strengthened is the use of pictures, both discrete (text-based) figures and visual drawings, as a problem solving domain. Through the presented example problems we aim to show the viability of the given problem domain for the teaching and learning of principles of programming, both from a procedural, process-based, as well as a concept, object-based, perspective. The given programming principles course has been taught since the academic year 1992/93 when the then newly established University of Cyprus (UCY) and its Computer Science Department (CS-UCY) admitted their first students. UCY was established by Law in 1989 as the first university in Cyprus. Its undergraduate students are being admitted through national examinations. For the CS-UCY undergraduates the entrance examinations include Mathematics, Informatics, Physics and Modern Greek. A notable achievement is that for the past five university entrances (2017-2021) CS-UCY's undergraduate programme has been ranked first amongst all undergraduate programmes in engineering and pure and applied sciences in Cyprus with respect to the number of candidate students including it as their top preference for university studies. Our undergraduate programme is continuously revised and modernized taking on board academic developments [8], as well as national developments at secondary school level. The latest ACM and IEEE proposals are presently under consideration.

The teaching and learning of fundamental principles of algorithmic thinking has been a core component and a standing challenge for Computer Science curricula [9, 11, 39, 47, 49] where key considerations include the usually diverse computer-based backgrounds and relevant abilities of incoming students. The currently observed diminishing interest of female students, in Europe and other parts of the world, in studying Computing [33] is also acquiring prominence calling for immediate and effective mitigation actions, in view of the fact that a large proportion of the current professions require relevant knowledge and skills, and this trend is predicted to be substantially increased in the foreseeable future, as new professions are perceived to be in even higher demand of computer specialists [10]. Unavoidably, computer programming classes give incoming students a first taste of what it is to follow in their curricula and as such it would not be an exaggeration to say that it is a critical factor regarding the choice of a student to stay or go, thus affecting in a major way the student retention rate. In this respect, as reported in the literature [42], female students may show higher interest towards visual problem domains in programming classes, in relation to other problem domains. We plan to investigate this conjecture in the context of our broader aim to encourage more females to follow university studies in computer science; in this paper only a cursory finding is presented, that bears some relation to what is reported in the literature.

The choice of a first programming language for university curricula in Computer Science and other related subjects has been a matter of continuous debate over the years [35, 38, 46, 48]. However, the key objective of our programming principles course (the CS131 course) has always been to familiarize students with the essential principles of programming and to help them develop their algorithmic thinking. As a highly practical problem solving course it entails the use of some high-level programming language, but learning whatever language is seen as a vehicle towards algorithmic problem solving. Initially the programming language used was the functional language Miranda [5, 15, 24]. Then it shifted to C/C++ [14] and since 2011 the course uses Java [44]. Regarding the course's visual feature which is the focus of this paper, in the initial period the high level aspects of Miranda made it easy to define abstract concepts like picture, figure, text or calendar, and to visualize them discretely or graphically, albeit in monochrome black and white, through some graphical front end. The use of Java in the current version of the course and in particular the use of graphical libraries and graphical objects enabled the strengthening of the visual feature, bringing in color, animation and even sound.

If we were to attempt a short historical review of the evolution of the course in terms of the programming languages and associated paradigms used, it can be said that a reason for abandoning the functional language Miranda in spite of the strengths of functional programming (abstraction, polymorphism, higher order functions, etc.) in supporting computational thinking and principled programming, was that students found recursion difficult to get into grips with. The lack of I/O and the declarative style of expression were sources of additional concern on the part of the students. Last but not least the fact that functional languages were not considered real application languages was a strong disincentive for students to learn such a language. Miranda was therefore swapped with C/C++. With C came the handling of pointers and associated memory management, which created new sources of difficulty for the students. In addition the coverage of such lower level features by necessity detracted attention from higher level algorithmic features. Furthermore, the coupling of C with C++, in practice did not go very smoothly. Most of the course was using C, while C++ was coming rather abruptly towards the end of the course to introduce objects, operator overloading, etc. Students were complaining that they had to learn two languages in one course. It was also the case that objects were not integrated in the course in a seamless fashion, in unison with the procedural aspects covered with C; instead objects appeared as an independent, stand-alone addition. A design weakness of the particular version of the course was that it did not deal with the same problems under the different paradigms to allow direct comparisons. Replacing the C/C++ combination by Java erased the above difficulties and so far we are happy with this choice. The students are happy with Java and the procedural/object-oriented synergy supported by this language serves well our objective of teaching principled programming and drawing direct comparisons between these two paradigms, as will be demonstrated in the example problems presented in Section 3.

We list abstraction, problem decomposition and synthesis, information hiding, reusability, modularity and extensibility as key principles of problem solving and algorithmic thinking. These principles are closely aligned with the advocated computational thinking techniques of problem decomposition, pattern recognition, pattern generalization and algorithm design [22, 31, 32, 37, 50]. Pattern generalization and pattern recognition are in fact abstraction and reusability. The messages on principled programming we try to convey to students are the following:

- Principled programming leads to algorithmic solutions that are more understandable, maintainable, reusable and extendible.
- Principled programming is about methodically building and testing (complex) structures, synergistically integrating two orthogonal concepts, clarity and detail, and demonstrating the purely benign side of "divide and conquer".
- Principled programming helps one to think abstractly when modeling processes and concepts.

Abstraction is critically significant not only to programming but to Computer Science in general as Kramer points out [28]. However, there is ample evidence that abstraction is difficult both to teach and to grasp "in the abstract" so to speak (see for example Hazzan's analysis on how students attempt to reduce abstraction [16]). From the pedagogical perspective, and as research has shown, abstraction should not be imposed on the students in a top down way as a kind of a commandment to follow blindly [40]; instead the students should be guided, by learning through their pitfalls as well, how to compose (or abduce) in a bottom up way, their own logistic path and appreciation to abstraction, when trying to model processes or concepts. More specifically, the construction of pictures, and the use of further visual aspects such as color, motion, and sound, can support the development of a constructive student perspective to problem solving [3].

It is important to clarify that the presented work is not about visual programming. Students write their programs textually and not graphically; it's the output of their programs which is in visual form. Our approach though is loosely related to the classical example domain of turtle graphics. Mini-languages, the development of which was seriously influenced by turtle graphics, and microworlds, can support computational thinking, often by non-Computer Science majors [7, 30]. Our focus is Computer Science majors, who should be able to design and build turtle objects and microworlds and not just use them, and also appreciate the multitude of algorithmic ways of producing the same visual output. Our initial inspiration in adopting the visual domain of pictures came from Bornat's classical book on programming from first principles [6].

Positive experience in using pictures in an introductory programming course for majors with media computation is reported by Simon et al. [45]; this work has similarities with our approach. Likewise, the discussed conceptual modeling [4] and components-first [19] approaches are rooted on object-oriented modeling and underlying principles (abstraction, modularity, reusability, information hiding, encapsulation) for teaching introductory programming; both approaches have strong commonalities with the overall pedagogy of the CS131 course.

The rest of the paper is organized as follows. Section 2 places the CS131 course in the broader context of the CS-UCY undergraduate curriculum and overviews some of its pedagogical features and student success rates. Section 3, vis-à-vis a number of illustrative example problems, drawn from the visual domain of discrete and graphical pictures, discusses the teaching and learning of the above principles of programming with respect to the procedural and object-oriented paradigms. Section 4, shows how students can get a glimpse of computational complexity in a visual, empirical way, and present their complexity measurements graphically. Finally, Section 5 concludes the discussion and points to some ongoing CS-UCY strategic endeavors.

## 2 The CS-UCY Programming Principles Course

UCY adopted the European Credit Transfer and Accumulation System (ECTS) [1, 2, 12, 23, 26, 27] and all its undergraduate curricula are four academic years of study (240 ETCS credits). The CS131 Programming Principles course is a semester 1 course and constitutes the starting node of no less than seven chains of twelve mandatory courses (see Figure 1).

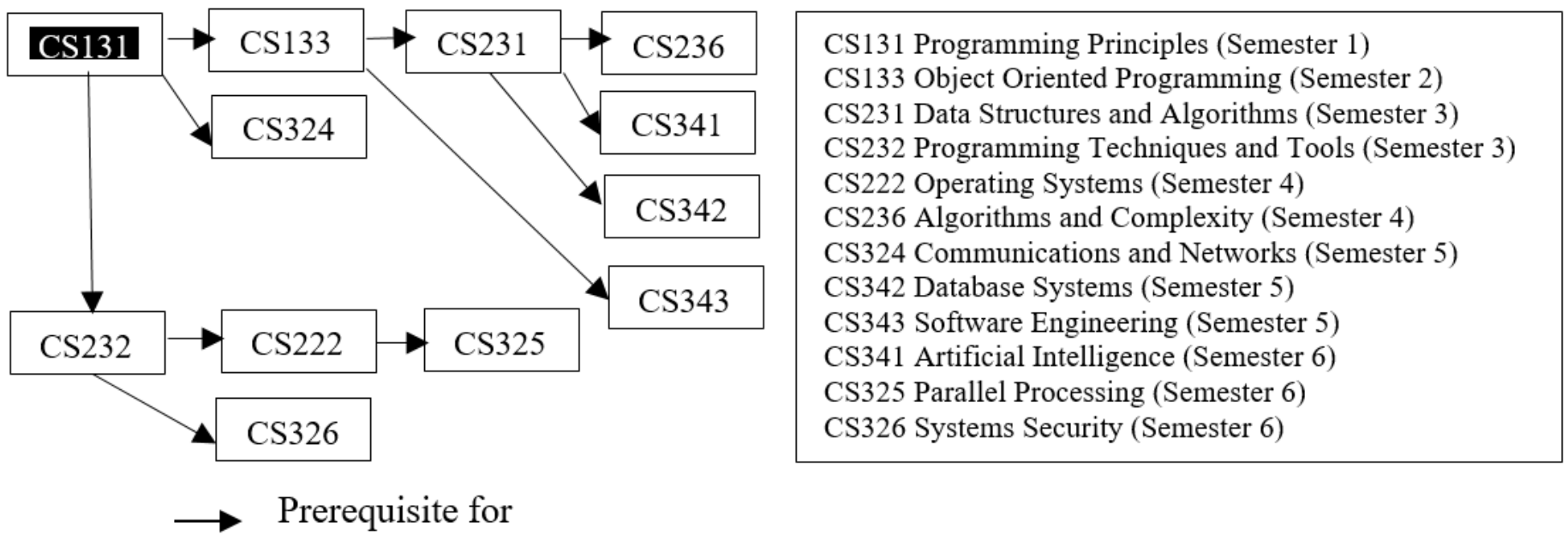

**Figure 1** The core position of the CS131 course in the CS-UCY undergraduate curriculum

### 2.1 Teaching and Learning Methods

The delivery of the CS131 course includes lectures, tutorials and supervised hands-on programming laboratories; the attendance in all these components is compulsory. Laboratory sessions are limited to 25 students while lectures and tutorials to 50 students. The material discussed in lectures and tutorials is closely coordinated with the hands-on exercises given in the laboratories. During lectures the various topics are presented in an interactive way, where students are encouraged to ask questions and express their views. Tutorials are strictly interactive sessions where students are expected not only to air their difficulties but also to initiate discussions between themselves. Some tutorials, usually towards the end of the course, are organized around case studies, where students are divided into groups of 3–5 and given a case study to discuss between them and then for each group to present its views to the other groups. In the laboratory sessions students are given coding exercises. During the run of the course, the students are given around 25 such coding exercises. The laboratory teachers present the exercises and provide guided assistance to the students.

### 2.2 Student Feedback and Evaluation Processes

Applying UCY's policy for continuous student evaluation/feedback, the course assessment includes a number of programming assignments, a mid-semester examination (or lately, due to the pandemic, two on-line quizzes) and a final examination. An illustrative example of an assignment from the problem domain of pictures is for students to copy computationally the art of Alison Turnbull[1] whose artistic creations include repetitive diagrams with small, random perturbations (see Figure 2). Students could exert their artistic skills and even to integrate into their creations motion and sound in addition to color. In the particular assignment, students were encouraged to approach the given picture construction problems both procedurally as well as through objects by defining the central structural elements of each diagram (wind mill, celestial body, etc.) as objects, and also to critically analyze the two approaches. Comparing and contrasting procedural and object-oriented solutions for the same problems, is an inborn aspect of the pedagogical methodology adopted in the CS131, as illustrated in Section 3.

Given that research has shown that feedback to the students can indeed support substantially student learning, e.g. [17, 36], particular attention is given throughout the CS131 course, to effective student feedback. In fact, the mid-semester examination serves exclusively the purpose of student feedback since if a student does better in the final examination, the mid semester examination does not count towards the overall course grade. Throughout the run of the course, the team of teachers (faculty and special teaching staff), are in close contact, not only for necessary coordination purposes, but also for detecting areas of general student difficulty or difficulties faced by specific groups of students, in an effort to try and address such difficulties as timely and effectively as possible.

### 2.3 Student Success Rates

The success rates of students (conversely their failure rates) are comparable to those reported in the literature for similar courses [47]. Table 1 gives the student numbers and success rates from the academic year 2016/17. The reading of these results should take on board the following:

- The numbers do not include those students that enrolled on the course but did not attend it.
- The fall semesters represent the regular delivery of the course and the spring semesters the repeat delivery. Thus during the fall semesters, the majority of students are doing the course for the first time, while during the spring semester the majority of students are doing the course for the second time.
- The Spring 2017 semester is an exception because owing to a change in legislation regarding the national service of male students, CS-UCY had an extraordinary

[1] https://alison-turnbull.com; visited on 11-January-2022.

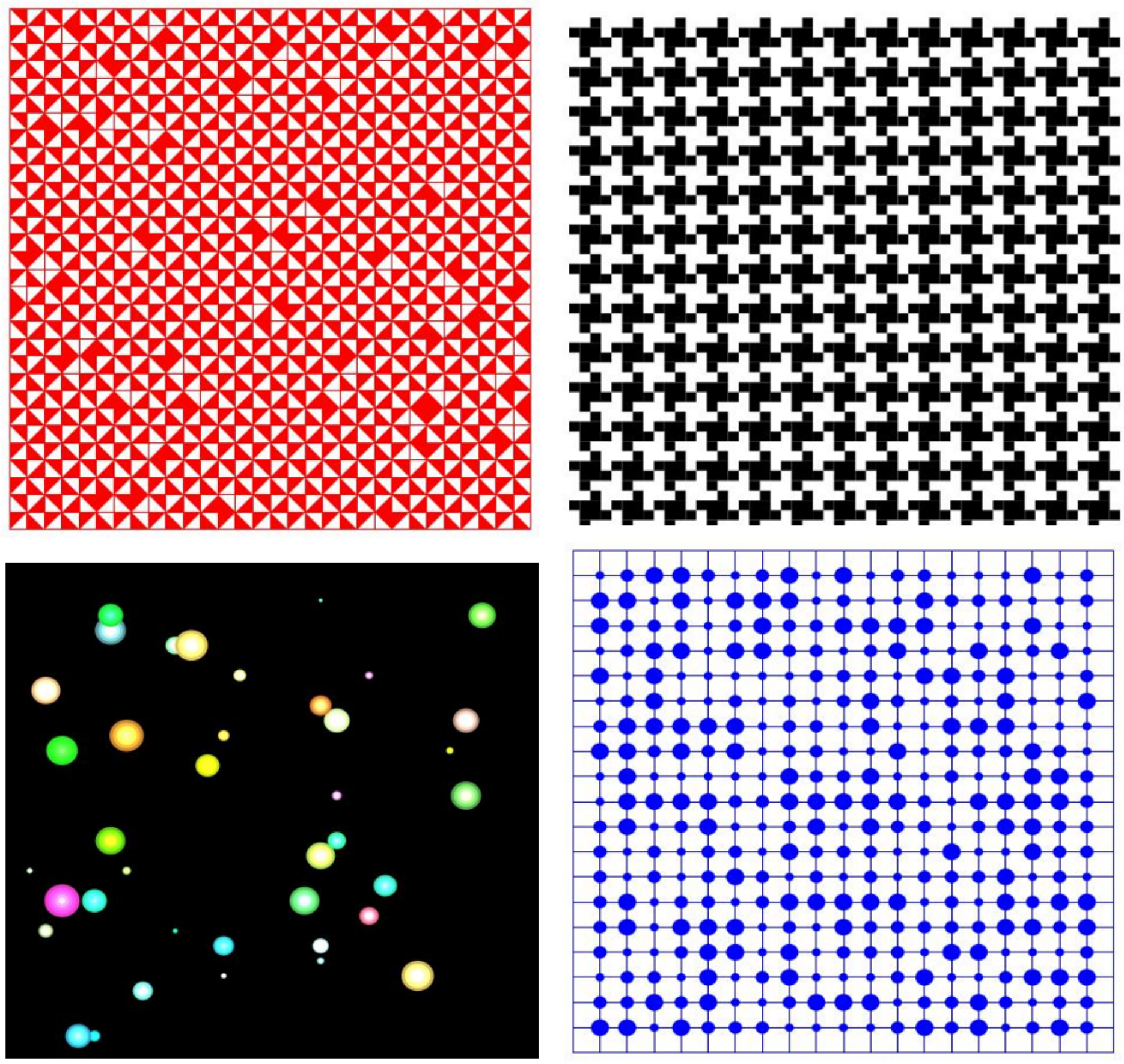

**Figure 2** Copying computationally the art of Alison Turnbull

admission of 39 male students during that semester, all doing the course for the first time.

- Over time, two groups of students appear to have particular difficulty with the course, attributed to the fact that they have no (or very sparse) background in computing from their secondary education. One of these groups is the students coming from secondary technical education (there is a special admission category for them) and the other group is that of overseas students who again are admitted on the basis of a special category and usually have largely diverse computing backgrounds. CS-UCY is in dialogue with the Ministry of Education regarding the mitigation of the issue concerning technical school graduates.
- The relatively higher success rate of the Spring 2019 semester, in spite of the fact that almost all students were repeating the course, may well be due to the fact that more extensive use than normally was made of the visual problem domain through graphical pictures. However, the higher success rate of the Fall 2020 semester is probably attributed to the Covid-19 situation, the fact that by necessity students had more time for their studies and the understandably higher leniency in the student evaluation.

**Table 1** Student success rates

| *Semester* | *#Passing* | *#Failing* | *#Total* | *%Success* |
|---|---|---|---|---|
| Fall 2016 | 59 | 20 | 79 | 74.68 |
| Spring 2017 | 32 | 25 | 57 | 56.14 |
| Fall 2017 | 66 | 25 | 91 | 72.53 |
| Spring 2018 | 20 | 12 | 32 | 62.50 |
| Fall 2018 | 65 | 30 | 95 | 68.42 |
| Spring 2019 | 28 | 3 | 31 | 90.32 |
| Fall 2019 | 64 | 17 | 81 | 79.01 |
| Spring 2020 | 19 | 11 | 30 | 63.33 |
| Fall 2020 | 83 | 9 | 92 | 90.22 |
| Fall 2021 | 87 | 19 | 106 | 82.08 |

## 3 Principles of Programming and the Visual Domain

We consider abstraction (both program and data abstraction), decomposition, synthesis, information hiding/encapsulation, reusability, modularity, extensibility and structuredness, as fundamental principles of programming and algorithmic problem solving. The given principles are strongly interrelated forming a tightly knit integral whole, and they are also closely aligned with the advocated computational thinking techniques, although under computational thinking abstraction is referred to as "pattern generalization", the ability to extract out unnecessary details and generalize those that are necessary to define a concept or idea in general terms, and reusability is coined as "pattern recognition", the ability to notice similarities, differences, properties, or trends in data. In this section, through a number of example problems taken from the CS131 course, we aim to demonstrate the viability of the visual domain of pictures with respect to the teaching/learning of the given principles. The code given is in Java.

```
$ java Pyramid 5 x
    x
   xxx
  xxxxx
 xxxxxxx
xxxxxxxxx

$ java Pyramid 7 o
      o
     ooo
    ooooo
   ooooooo
  ooooooooo
 ooooooooooo
ooooooooooooo

$ java Pyramid 9 p
        p
       ppp
      ppppp
     ppppppp
    ppppppppp
   ppppppppppp
  ppppppppppppp
 ppppppppppppppp
ppppppppppppppppp
```

**Figure 3** Drawing pyramids

■ **Listing 1** First version of the Pyramid program using character print operations

```
public class Pyramid {
 public static void main (String[] args){
   int h = Integer.parseInt(args[0]);
   char c = args[1].charAt(0);

   for (int r = 1; r <= h; r++){
      for (int j = 1; j <= h-r; j++) System.out.print(' ');
      for (int j = 1; j <= 2*r-1; j++) System.out.print(c);
      System.out.print('\n');
   }
 }
}
```

### 3.1 The Pyramid Problem

An early course problem is the drawing of pyramids in a discrete fashion. The students are asked to write a program that gets from the command line a number (h) and a character (c) and constructs a pyramid as illustrated in Figure 3. First the students are asked to glean out the structure of the given pyramids. They can see straightaway that such a pyramid consists of $h$ rows, where the given character (c) appears 1, 3, 5, .., $2h-1$, times from the top (number 1) row to the bottom (number $h$) row. It takes them a bit longer to generalize this by saying that each row $r$, consists of two segments, a first segment of length $h-r$ made up from spaces and a second segment of length $2r-1$ made up from the c character. Once they discover the structure of the pyramid picture, the first attempt at writing the program comes quite naturally to the program shown in Listing 1. This program unrolls into a sequence of individual character print operations.

Clearly at this stage, the students had viewed the pyramid at the micro level of its individual component characters, and not in its totality. To help them switch to the macro level view, they are asked to think of a pyramid in totality as a string, and to replace the print character operations with string concatenation operations, thus resulting in the modified *Pyramid* program shown in Listing 2.

■ **Listing 2** Second version of the Pyramid program using string concatenations

```
public class Pyramid {
 public static void main (String[] args){
   int h = Integer.parseInt(args[0]);
   char c = args[1].charAt(0);
   String s = "";
   for (int r = 1; r <= h; r++){
      for (int j = 1; j <= h-r; j++) s += ' ';
      for (int j = 1; j <= 2*r-1; j++) s += c;
      s +='\n';
   }
  System.out.print(s);
 }
}
```

**Listing 3** Modular version of the Pyramid program

```
public class Pyramid {
 public static String line (int len, char c){
    String s = "";
    for (int i = 1; i <= len; i++) s += c;
    return s; }

 public static String pyramidRow (int h, int r, char c){
   return line(h-r, ' ') + line(2*r - 1, c) + line(h-r, ' ');}

 public static String drawPyramid (int h, char c){
   String s = "";
   for (int i = 1; i <= h; i++) s += pyramidRow(h,i,c) + "\n";
   return s;}

 public static void main (String[] args){
   int h = Integer.parseInt(args[0]);
   char c = args[1].charAt(0);
   System.out.println(drawPyramid(h,c));}
}
```

These simple code changes resulting in the construction of the given discrete picture, in totality as a string, give an interesting insight to the students gradually leading them to the realization that any discrete picture could indeed be constructed as such. Still the *Pyramid* program comes with no modularity. The students tend to have difficulty in identifying some higher level constructs leading to useful abstractions for their pyramid drawing.

When a simple method for reversing a string was included in the program and used to apparently create, in a single step, the image of the pyramid upside down, the students surprisingly observed that the reversed image of the pyramid was not in fact an upside down pyramid, but a right angle triangle. The analysis of the cause of this result, revealed that the symmetry could be achieved in the reversed (upside down) image of a pyramid, if a third segment were to be concatenated to each row, the same as the first segment. The students could now easily construct, in simple steps, vertical combinations of pyramids in up or down orientations, but had difficulty in figuring out how horizontal arrangements of pyramids could be drawn. The next step was to reformulate the *Pyramid* program in a modular fashion as shown in Listing 3 and through this exercise to see a simple hierarchical program structure, where the basic abstraction construct is method *line* for composing a same character string of a given character and length. But more importantly there are now reusable methods, that in fact provide the means for drawing horizontally arranged pyramids, namely method *pyramidRow* that separates the generation of a given row of a pyramid, as a distinct entity. More functionality is thus gained through the aforementioned decomposition into separately defined, reusable modules. It is also important to note that once students come to view parts of pyramids, such as their rows, or the pyramids themselves in totality, as strings which can be given as input to methods, or produced as output by methods, their computational thinking has made a big step from the initial micro level view of a pyramid as a sequence of print character operations, towards the direction of principled algorithmic thinking and the construction of reusable abstractions.

This discussion paved the way towards the subsequent definition of a pyramid object and more generally of a discrete picture object, *DiscretePic*, that provides ample functionality in a modular way (different ways of constructing instances of discrete pictures, framing pictures, joining them vertically or horizontally, scaling them, overlaying or clipping them, exporting them externally as strings, storing them, etc.). Moreover, by applying information hiding (encapsulation) the students can experiment with different ways of implementing the *DiscretePic* object, e.g. as a two dimensional table of characters or as a table of strings, without affecting the object's public view. Figure 4 gives some example instances of object *DiscretePic* and Figure 5 gives the August 2019 instance of another object type, *CalendricMonth* where the letters in the *CalendricMonth* objects are in fact *DiscretePic* objects appropriately scaled to look bold and positioned as a horizontal banner.

```
+---------------++---------------+
|+-------------+||+-------------+|
||      *      ||||*************||
||     ***     |||| *********** ||
||    *****    ||||  *********  ||
||   *******   ||||   *******   ||
||  *********  ||||    *****    ||
|| *********** ||||     ***     ||
||*************||||      *      ||
|+-------------+||+-------------+|
+---------------++---------------+
```

```
      *
     ***
****ooooo****
 **ooooooo**
  ooooooooo
 **ooooooo**
****ooooo****
     ***
      *

      *
     ***
****ooooo****
 **ooooooo**
  ooooooooo
 **ooooooo**
****ooooo****
     ***
      *
```

```
              o
             ooo
            ooooo
           ooooooo
          ooooooooo
ooooooooo***********ooooooooo
 ooooooo*************ooooooo
  ooooo***************ooooo
   ooo*****************ooo
    o*******************o
    o*******************o
   ooo*****************ooo
  ooooo***************ooooo
 ooooooo*************ooooooo
ooooooooo***********ooooooooo
          ooooooooo
           ooooooo
            ooooo
             ooo
              o
```

**Figure 4** Example instances of object DiscretePic

### 3.2 The Text Problem

Text processing provides fertile ground for student problem solving and coding activities that can be interestingly approached as discrete figure drawing. If the appropriate "pattern generalizations" can be made, seemingly different problems can converge to the same abstraction enabling the application of a generalized solution.

A recent text processing programming activity, given to the students as an online programming task, was to write a program (*TabularText*) to process the input of some text file and to present it in a tabular form. Figure 6 gives such an example text contained in file sample.txt and Figure 7 gives the output of the program for the given example text. The program takes as input the size of each column in terms of width (number of characters) and height (number of lines), both of which have the value 20 in this example run. For this run of the program, three columns of the specified size were needed for the conversion of the given text into tabular form. Students appreciate that the content of a text file is a sequence of characters that can be read

```
  ******    **        **    ******    **        **    ******    **********
  ******    **        **    ******    **        **    ******    **********
**      **  **        **  **      **  **        **  **              **
**      **  **        **  **      **  **        **  **              **
**********  **        **  **          **        **    ******        **
**********  **        **  **          **        **    ******        **
**      **  **        **  **    ****  **        **          **      **
**      **  **        **  **    ****  **        **          **      **
**      **    ********      ******      ********      ******        **
**      **    ********      ******      ********      ******        **

2019

 Monday     Tuesday    Wednesday  Thursday   Friday     Saturday   Sunday
+----------+----------+----------+----------+----------+----------+----------+
|          |          |          | 1        | 2        | 3        | 4        |
|          |          |          |          |          |          |          |
|          |          |          |          |          |          |          |
|          |          |          |          |          |          |          |
|          |          |          |          |          |          |          |
+----------+----------+----------+----------+----------+----------+----------+
| 5        | 6        | 7        | 8        | 9        |10        |11        |
|          |          |          |          |          |          |          |
|          |          |          |          |          |          |          |
|          |          |          |          |          |          |          |
|          |          |          |          |          |          |          |
+----------+----------+----------+----------+----------+----------+----------+
|12        |13        |14        |15        |16        |17        |18        |
|          |          |          |          |          |          |          |
|          |          |          |          |          |          |          |
|          |          |          |          |          |          |          |
|          |          |          |          |          |          |          |
+----------+----------+----------+----------+----------+----------+----------+
|19        |20        |21        |22        |23        |24        |25        |
|          |          |          |          |          |          |          |
|          |          |          |          |          |          |          |
|          |          |          |          |          |          |          |
|          |          |          |          |          |          |          |
+----------+----------+----------+----------+----------+----------+----------+
|26        |27        |28        |29        |30        |31        |          |
|          |          |          |          |          |          |          |
|          |          |          |          |          |          |          |
|          |          |          |          |          |          |          |
|          |          |          |          |          |          |          |
+----------+----------+----------+----------+----------+----------+----------+
```

**Figure 5** The August 2019 instance of object CalendricMonth

as individual characters, or as words (separated by white space) or as lines (separated by new line characters) or in totality as a string.

Listing 4 gives the code produced by one of the students (the code uses standard input methods from the *StdIn* library of the course textbook [44]). The particular student saw the analogy with the pyramid problem and more specifically the analogy of this problem with the horizontal placement of a number of pyramids of the same size, irrespective of their orientations, upwards or downwards, since each pyramid occupies the same rectangular space, just like the text columns that needed to be placed next to each other. In order to make the required justification to the left of each column, each column's lines must be of the same width. This important sub-problem was identified by the particular student, who coded it as a separate method (*fillLine*) that extends a string with the relevant number of space characters. No other modularity was used in this student's solution, but method *fillLine* abstracts crucial functionality.

```
Computer programming often shortened to programming
is a process that leads from an original formulation of a
computing problem to executable computer
programs. Programming involves activities such as analysis,
developing understanding, generating algorithms,
verification of requirements of algorithms including their
correctness and resources consumption and implementation
of algorithms in a target programming language.
Source code is written in one or more programming languages.
The purpose of programming is to find a sequence of
instructions that will automate performing a specific
task or solving a given problem. The process of programming
thus often requires expertise in many different subjects
including knowledge of the application domain
specialized algorithms and formal logic.
```

**Figure 6** Example text contained in file sample.txt

```
$java TabularText 20 20 < sample.txt

Computer programming  consumption and       expertise in many
often shortened to    implementation of     different subjects
programming is a      algorithms in a       including knowledge
process that leads    target programming    of the application
from an original      language. Source      domain specialized
formulation of a      code is written in    algorithms and
computing problem to  one or more           formal logic.
executable computer   programming
programs.             languages. The
Programming involves  purpose of
activities such as    programming is to
analysis, developing  find a sequence of
understanding,        instructions that
generating            will automate
algorithms,           performing a
verification of       specific task or
requirements of       solving a given
algorithms including  problem. The process
their correctness     of programming thus
and resources         often requires
```

**Figure 7** Example run of the TabularText program

Another interesting feature of this solution is the use of a one-dimensional table of strings, whose size is given by the number of rows of each column, and each of its entries eventually holds the lines corresponding to each row of the columns, concatenated together, and separated with a specified gap of two spaces. Thus as soon as the lines forming a column are incorporated in the table entries and there is still text to be processed, the iteration reverts back to the first element of the table and starts composing and joining the lines of the next column. Eventually all elements of the table are joined together into a single string representing the original text in its new tabular form. This string is then displayed. There were other interesting student solutions for this programming task, but these lacked the simplicity of the above solution and by and large had no modularity. In general, students need time to gradually acquire the computational skill of decomposing a problem into sub-problems, leading to separately defined (reusable) modules.

**Listing 4** TabularText program written by a student

```
public class TabularText {
  private static String fillLine (String s, int w){
    while (s.length() < w) s += ' ';
    return s;}

  public static void main (String[] args){
    int w = Integer.parseInt(args[0]), h = Integer.parseInt(args[1]);
    int gap = 2;
    String[] lines = new String[h];
    for (int i = 0; i < lines.length; i++) lines[i] = "";
    String line = null;
    int count = 0;
    while(!StdIn.isEmpty()){
      String word = StdIn.readString();
      if (line == null) line = word;
      else if ((line + ' ' + word).length() <= w) line += ' ' + word;
      else {if (count == h) count = 0;
            lines[count] += fillLine(line, w+gap); line = word; count++;}
    }
    if (count == h) {count = 0;}
    lines[count] += fillLine(line,w+gap);
    String text = "";
    for (String ls : lines) {if (!ls.equals("")) {text += ls + "\n";}}
    System.out.println(text);}
}
```

The common student pitfalls in this particular programming task were the inability to achieve the required column justification, although the columns were correctly identified and placed next to each other, or failing to place the columns in a horizontal fashion, but rather in a vertical, or even diagonal fashion. The latter pitfall was the consequence of either not using any data structure for storing the column lines, or failing to achieve the joining of the corresponding lines for each column row to get the horizontal arrangement.

The above programming task was done by 101 students, 17 female students and 84 male students (unfortunately the percentage of female students has decreased substantially over the last few years). In total 15 students, i.e. 14.85 %, produced excellent solutions, scoring at least 90 out of 100. It is interesting to note that in these 15 students, there were 4 female students and 11 male students, and thus the respective percentages of excellent performance amongst the female and male students were 23.53 % and 13.1 %. This small piece of evidence is consistent with research results regarding the interest of female students in problems with visual output [42].

As mentioned above, students need time to incrementally develop the computational skill of decomposing a problem into sub-problems, dealing separately with each sub-problem and then synthesising them into the overall problem solution. By trying to handle a problem in its entire complexity, and putting everything into a single module, students tend to end up producing unnecessarily complex solutions prone to logical errors that are difficult to trace and correct.

**Listing 5** Program ModularTabularText

```
public class ModularTabularText {
 private static String fillLine (String s, int w){
   while (s.length() < w) s += ' ';
   return s;}

 public static String[] getLines(int w, int h){
   int gap = 2;
   String[] lines = new String[h];
   for (int i = 0; i < lines.length; i++) lines[i] = "";
   String line = null;
   int count = 0;
   while(!StdIn.isEmpty()){
     String word = StdIn.readString();
     if (line == null) {line = word;}
     else if ((line + ' ' + word).length() <= w) {line += ' ' + word;}
     else {if (count == h) {count = 0;}
           lines[count] += fillLine(line, w+gap); line = word; count++;}
   }
   if (count == h) {count = 0;}
   lines[count] += fillLine(line,w+gap);
   return lines;}

 public static String joinLines(String[] lines){
   String text = "";
   for (String ls : lines) if (!ls.equals("")) {text += ls + "\n";}
   return text; }

 public static void main (String[] args){
   int w = Integer.parseInt(args[0]), h = Integer.parseInt(args[1]);
   System.out.println(joinLines(getLines(w,h)));}
}
```

During the follow up tutorial session on the above programming task, a modular solution was produced with the help of the students based on the aforementioned student solution. This is shown in Listing 5 as program *ModularTabularText*. In addition to the *fillLine* method, two other methods, *getLines* and *joinLines* are included, that abstract the handling of two other sub-problems of the particular problem. Once a modular program was defined by separating key functions, students were then encouraged to capitalize on the given program abstractions, turning them into an abstract data type, namely object *TextObj*, representing textual objects. The functionality of this object was sketched, leading to the class definition given in Listing 6.

Objects are by nature (complex) modular entities. Reference methods, *getWords* and *tabulate* are included in the public view of object *TextObj*. Extensibility is readily supported as additional functionality through new reference methods, or variant definitions of existing reference methods through function overloading, can be easily included in an object's public view. In fact, students were asked to extend the *TextObj* class by including reference methods for turning its low-case letter characters to capitals, or centralizing its lines, or justifying its lines to the right, etc., in order to appreciate the easiness of such modular extensions.

**Listing 6** Defining object TextObj

```
public class TextObj{
 private String content;

 public TextObj (String text){content = text;}

 private static String fillLine (String s, int w){. . .}

 private static boolean whiteSpace(char c){
   return c == ' ' || c == '\n' || c == '\t'; }

 private static String[] addWord(String[] ws, String word){. . .}

 private static String joinLines(String[] lines){. . .}

 public String[] getWords(){
   String[] ws = null; String word = "";
   boolean beg = false;
   for (int i = 0; i < content.length(); i++){
     if (beg && whiteSpace(content.charAt(i))) {
               ws = addWord(ws, word); word = ""; beg = false;}
     else if (!beg && !whiteSpace(content.charAt(i))){
                 word += content.charAt(i); beg = true;}
     else if (beg && !whiteSpace(content.charAt(i)))
                   word += content.charAt(i);}
   if (!word.equals("")) ws = addWord(ws, word);
   return ws;}

 public void tabulate (int w, int h){
   int gap = 2;
   String[] lines = new String[h];
   for (int i = 0; i < lines.length; i++) lines[i] = "";
   String line = null;
   int count = 0;
   for (String word: getWords()){
     if (line == null) line = word;
     else if ((line + ' ' + word).length() <= w) {line += ' ' + word;}
     else {if (count == h) {count = 0;}
             lines[count] += fillLine(line, w+gap); line = word; count++;}
   }
   if (count == h) count = 0;
   lines[count] += fillLine(line,w+gap);
   content = joinLines(lines); }

 public String toString(){return content;}
}
```

As it can be seen, both in the pyramid and the text problems, the initial attempt was to get a solution, albeit in a procedural, non-modular and thus relatively unstructured way. The next attempt was to improve the procedural solution by adding explicit structure through separately defined modules, and eventually to turn this to an object-based solution with clear interfaces for the provided functionality.

**Listing 7** A procedural solution to the moving circle problem

```
public class MovingCircle {// procedural version
 public static void main (String[] args){
   double x = Double.parseDouble(args[0]); // x-coord of centre of fixed circle
   double y = Double.parseDouble(args[1]); // y-coord of centre of fixed circle
   double R = Double.parseDouble(args[2]); // radius of fixed circle
   double r = Double.parseDouble(args[3]); // radius of moving circle
   StdDraw.setXscale(0.0,100.0);
   StdDraw.setYscale(0.0,100.0);
   StdDraw.setPenRadius(0.01);
   while (true){
     double angle = Math.random()*360.0; // random angle from [0..360]
     double n = Math.random()*R; // random distance from [0 .. R]
     double x1 = n * Math.cos(Math.toRadians(angle)) + x;
     double y1 = n * Math.sin(Math.toRadians(angle)) + y;
     double d = Math.sqrt((x-x1)*(x-x1) + (y-y1)*(y-y1));
     if (d+r <= R) {
        StdDraw.clear(StdDraw.WHITE);
        StdDraw.setPenColor(StdDraw.BLACK);
        StdDraw.circle(x,y,R);
        StdDraw.setPenColor(StdDraw.RED);
        StdDraw.circle(x1,y1,r);
        StdDraw.show(200); // freeze for 200 microseconds
     }
   }
 }
```

An intended learning objective, therefore, is to guide the students, through interactive discussions and collaborative problem solving, to compose, compare and contrast the procedural and object-oriented solutions of the same problem, as well as the modular and non-modular solutions. As already mentioned such comparisons are an inborn aspect of the pedagogical methodology of the CS131 course.

### 3.3 The Moving Circle Problem

Another example problem, this time involving a graphical solution, is the moving circle problem, where a smaller circle continuously moves within a bigger fixed circle. A snap shot of the visual solution of this problem, without the animation, is shown in Figure 8, and Listing 7 gives the initial, procedural version, of program *MovingCircle*. The program takes as input the coordinates of the centre of the larger fixed circle ($x$ and $y$), the radius of this circle ($R$) and the radius of the smaller moving circle ($r$) and produces a visual, animated solution. For this drawing the program uses methods from the *StdDraw* library of the course textbook [44]. The *StdDraw* library, that includes methods for plotting points and drawing line segments, or filled or not filled, squares, circles, and more generally polygons, is extensively used in the course. The use of this library and other libraries given in the course textbook, as well as standard libraries of Java, such as the *Math* library which is also utilized in this problem, helps the students to gain further insights on the importance of abstraction, modularity, information hiding and reusability. They simply take 'off the shelf', ready-made modules providing important functionality that can be readily deployed just by understanding the modules' interfaces and functionality and nothing about their implementation; they are treated as useful 'black box' aids for configuring solutions to different problems.

The principal algorithmic structure of the *MovingCircle* program is that of an infinite loop, for continuously selecting, in a random fashion, the coordinates of a point within the fixed circle to form the centre of the small circle provided that the small circle will remain within the big circle. Once these coordinates are selected, the previous picture of the two circles is erased, the new picture, in which only the small circle changes position, is redrawn and frozen for 200 microseconds, thus giving the impression of movement. This is a simple program, generating a simple animated picture, that nonetheless triggers the interest of students that tend to approach this problem quite enthusiastically.

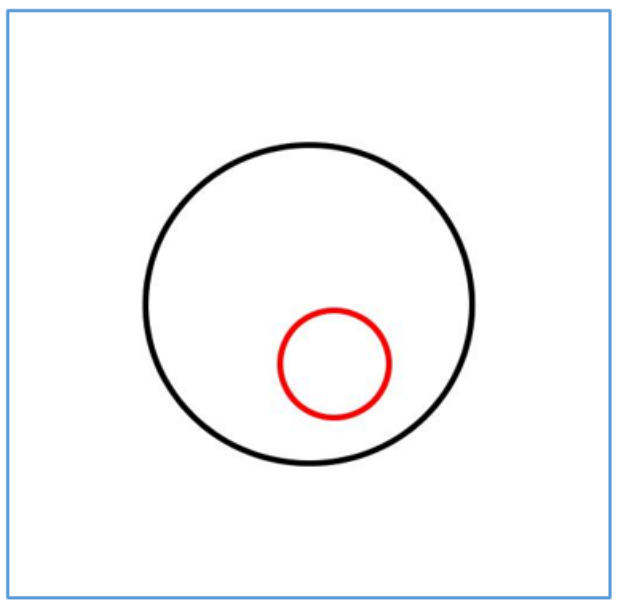

**Figure 8** The moving circle problem

The follow up discussion around this problem is for students to identify the entities involved and to try and model them, not just for the specific needs of the given problem, but to ascribe broader functionality to them, for more general use. This discussion results in the definitions of objects *Point* and *Circle*, as shown in Listings 8 and 9. Modularity, information hiding or rather encapsulation, reusability and extensibility are intrinsically related to objects and as such these principles are relatively easier to convey to students when objects are discussed. The *Point* object represents fixed points in a Cartesian space. Moreover, points can be produced from a given point through scaling, shifting or rotating operations. A point can be plotted, it can form a line segment with another point, or its distance with another point can be calculated. The centre of a *Circle* object is a *Point* object. By changing its centre to another point, a circle moves. Furthermore, a circle's area and periphery can be calculated, it can be decided whether a circle contains a point or another circle (thus exploiting function overloading) or whether it intersects or is disjoint with another circle. Finally a *Circle* can be drawn in its own color (thus also making use of standard object *Color*). The functionality bestowed to these two objects is independent of any of their potential uses and more importantly it can be easily extended without affecting the existing functionality. This reusability message is strongly conveyed to the students. Objects *Point* and *Circle*, provide sufficient ammunition to the students for converting the procedural version of the *MovingCircle* program to an object-oriented version, and this is what they are asked to do next, and to critically compare and contrast the two versions. An object-oriented version of the program is given in Listing 10. Students are also prompted to modify these programs to solve a number of variants of this problem, e.g. when more than one smaller circles, continuously and randomly move within the bigger fixed circle, either independently, or in a nested fashion. In addition to the entertaining aspect of producing visual, animated, solutions to this problem, the students are asked to see which approach facilitated better the modifications, the procedural, or the object-oriented one, especially if the number of the smaller circles is substantially increased.

**Listing 8** Defining object Point

```
public class Point {
 private final double Xcoord;
 private final double Ycoord;

 public Point (double x, double y){
   Xcoord = x; Ycoord = y;}

 public double getX(){return Xcoord;}
 public double getY(){return Ycoord;}

 public Point scale (double a, double b){
   double x = a * Xcoord;
   double y = b * Ycoord;
   return new Point(x,y);}

 public Point shift (double a, double b){
  double x = a + Xcoord;
  double y = b + Ycoord;
  return new Point(x,y);}

 public Point rotate (double angle){
   double theta = angle * Math.PI/180;
   double x = Xcoord * Math.cos(theta) - Ycoord * Math.sin(theta);
   double y = Xcoord * Math.sin(theta) + Ycoord * Math.cos(theta);
   return new Point(x,y);}

 public double distanceTo (Point p){
  double dx = Xcoord - p.Xcoord;
  double dy = Ycoord - p.Ycoord;
  return Math.sqrt(dx*dx + dy*dy);}

 public void Line (Point p){
   StdDraw.line(Xcoord, Ycoord, p.Xcoord, p.Ycoord);}

 public void plot(){
   StdDraw.point(Xcoord, Ycoord);}

 public String toString (){
  return "Point at (" + Xcoord + "," + Ycoord + ")";}
}
```

**Listing 9** Defining object Circle

```
import java.awt.Color;
public class Circle {
 private final double radius;
 private Point center;
 private final Color color;

 public Circle (double radius, Point center, Color color){
     this.radius = radius;
     this.center = center;
     this.color = color;}

 public boolean contains(Point p) {
     return p.distanceTo(center) <= radius;}

 public double area() {
     return Math.PI * radius * radius;}

 public double periphery(){
     return Math.PI * 2 * radius;}

 public boolean intersects(Circle c) {
     return center.distanceTo(c.center) <= radius + c.radius;}

 public boolean disjoint (Circle c){
     return !intersects(c);}

 public boolean contains (Circle c){
     return center.distanceTo(c.center) + c.radius <= radius; }

 public void move (Point p){
    center = p;}

 public void draw(){
   StdDraw.setPenColor(color);
   StdDraw.circle(center.getX(), center.getY(), radius);}
}
```

### 3.4 The Dragon Curve Problem

The last problem presented, is the graphical drawing of one of the well known fractals, namely the dragon curve (see Figure 9 for an example of this curve of depth 10). Fractals are recursive drawings and as such through this problem, recursion is also demonstrated, in an object-based fashion (a procedural, recursive, version is also discussed with the students, but not shown here). In addition, the reusability potential of the *Point* object is also demonstrated, as the particular object is now deployed for a completely different problem. The resulting object definition, *DragonCurve*, is shown in Listing 11. As it can be seen from this definition, internally an instance of object *DragonCurve* is characterised by a line segment of unit length defined by a pair of *Point* instances (start and finish), its number of levels or its depth ($n$) and its color.

An *DragonCurve* instance of depth zero (0) is just the given line segment drawn in the given color. An instance of depth one (1) consists of four line segments that are obtained from the initial line segment through relevant transformations, one being the scaling by half of the initial line segment. This process is recursively applied until the depth is reduced to 0 and each time four new line segments, half the size, are obtained from each of the line segments of the previous (depth) level. Recursion is used in the definition of reference method *draw* that actually has two variants; the public variant that has no arguments, and the private variant that has as input an integer ($n$) denoting a depth and two *Points* denoting a line segment. For a dragon curve drawing operation, the public variant of *draw* is initially called, and this in turn calls its private variant; the latter calls itself recursively four times until the basic cases of zero depths are reached. It is noted that yet again function overloading, a useful feature of Java, is used.

The complexity of the dragon drawing problem, in terms of the number of line segments that need to be produced and drawn, as a function of a curve's depth, is one aspect discussed with the students. For example, the dragon curve of Figure 9 consists of 1048576 line segments ($4^{10}$). These are tiny line segments, more like points than lines, since their length is $1/2^{10}$ (the initial line segment is of unit length (1)). The students can experiment with creating and drawing instances of the *DragonCurve* object, of different depths, and observe in real time the drawing of the individual line segments. For the higher depths, all the space is eventually filled up. Students can also experiment with instances having different start and finish points, although such distinctions are only noticeable for smaller depth instances of the drawing. For higher depths, each drawing converges to the same graphical picture.

**Listing 10** Object-oriented solution of the moving circle problem utilizing objects Point and Circle

```
public class MovingCircle { // using objects Circle and Point
 public static void main (String[] args){
   double x = Double.parseDouble(args[0]); // x-coord of centre of fixed circle
   double y = Double.parseDouble(args[1]); // y-coord of centre of fixed circle
   double R = Double.parseDouble(args[2]); // radius of fixed circle
   double r = Double.parseDouble(args[3]); // radius of moving circle
   StdDraw.setXscale(0.0,100.0);
   StdDraw.setYscale(0.0,100.0);
   StdDraw.setPenRadius(0.01);
   Circle c1 = new Circle(R, new Point(x,y), StdDraw.BLACK);
   while (true){
     double x1 = Math.random()*2*R + x - R; // random number from [x-R .. x+R]
     double y1 = Math.random()*2*R + y - R; // random number from [y-R .. y+R]
     Circle c2 = new Circle(r, new Point(x1,y1), StdDraw.RED);
     if (c1.contains(c2)) {
        StdDraw.clear(StdDraw.WHITE);
        c1.draw();
        c2.draw();
        StdDraw.show(200); } // freeze for 200 microseconds
   }
  }
}
```

**Listing 11** Object DragonCurve

```
import java.awt.Color;
public class DragonCurve{
  private Point start, finish;
  private int n; // the levels
  private Color color;

  public DragonCurve(Point start, Point finish, int n, Color color){
    this.start = start;
    this.finish = finish;
    this.n = n;
    this.color = color; }

  public void draw(){
    StdDraw.setPenColor(color);
    draw(n, start, finish);}

  private void draw (int n, Point p1, Point p2){
     if (n==0){p1.Line(p2); return;}
     draw(n-1,p1.scale(0.5,0.5).rotate(90.0),
              p2.scale(0.5,0.5).rotate(90.0));
     draw(n-1,p1.scale(0.5,0.5).rotate(180.0).shift(0.5,0.5),
              p2.scale(0.5,0.5).rotate(180.0).shift(0.5,0.5));
     draw(n-1,p1.scale(0.5,0.5).rotate(-90.0).shift(0.5,0.5),
              p2.scale(0.5,0.5).rotate(-90.0).shift(0.5,0.5));
     draw(n-1,p1.scale(0.5,0.5).rotate(180.0).shift(1.0,0.0),
              p2.scale(0.5,0.5).rotate(180.0).shift(1.0,0.0));}
}
```

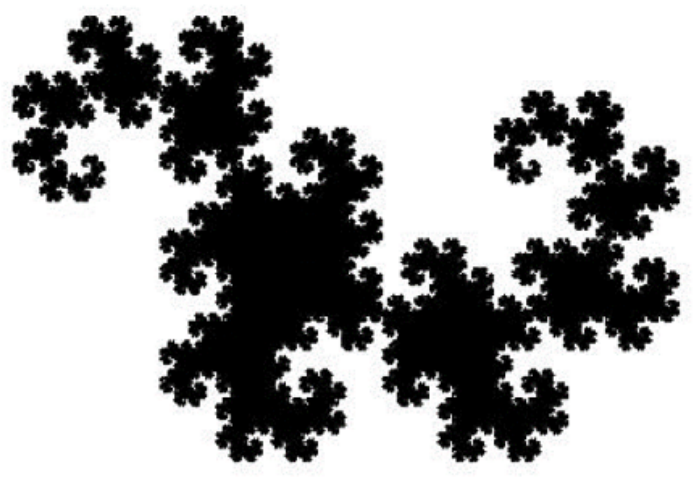

**Figure 9** The dragon curve

This example problem, also gives the opportunity to illustrate and discuss how reference methods can be chained together, and how such chains are evaluated (from left to right) and that each evaluation generates a new object instance, that will proceed to activate the reference method which is the next one in the chain, until the end result is produced. In this example the chains are pieced together with reference methods of the *Point* object (scale, shift, rotate) and thus all the intermediate object instances generated are *Point* instances. Students need to appreciate, though, and they are shown relevant examples, that generally speaking such chains or pipes of reference methods can yield intermediate object instances of different object types. A challenge given to the students is to calculate the number of *Point* instances generated for the drawing of a dragon curve of a given depth. Interesting exchanges can accrue from these complexity findings, that also impinge on recursive versus iterative processing (e.g. is recursion more natural to use for the drawing of a recursive picture?), procedural versus object-based approaches, as well as simplicity and transparency of code. It is true that students tend to be (justifiably) concerned about the large number of intermediate *Point* instances created en route to the drawing of a dragon curve. It is also true that such concerns are not necessarily erased, simply by pointing out to them that such instances, after they would have run their course, are dealt with by the garbage-collector, working in a self-driven way behind the scenes. In this example problem, students appear to appreciate that the object-based solution, supersedes the procedural one, in that it is more natural in expressing the computation involved. Yet another issue, is to contrast a chain of reference methods (arising in this and other examples) against a nesting of static method calls, evaluated from the innermost to the outermost method call, where the output of one method is channeled as the input of the next method. This discussion once more revolves around the principle of modularity and the encouragement of students to apply this principle, both statically yielding useful program abstractions, and with respect to object reference methods, where modularity should be seen as an inborn feature of objects.

### 3.5 Other Visual Examples

Other graphical creations, drawn using methods from the *StdDraw* library as well as trigonometric functions, involve colorful drawings constructed by simple repetitive operations. In one exercise students are asked to convert one picture (the star picture) through a single transformation and appropriate color changes to another picture (the sun picture); both pictures are shown in Figure 10. The aim of this exercise was for students to appreciate that a well-designed algorithmic method can be easily modified to obtain another method, with impressively distinct, from the visual perspective, output.

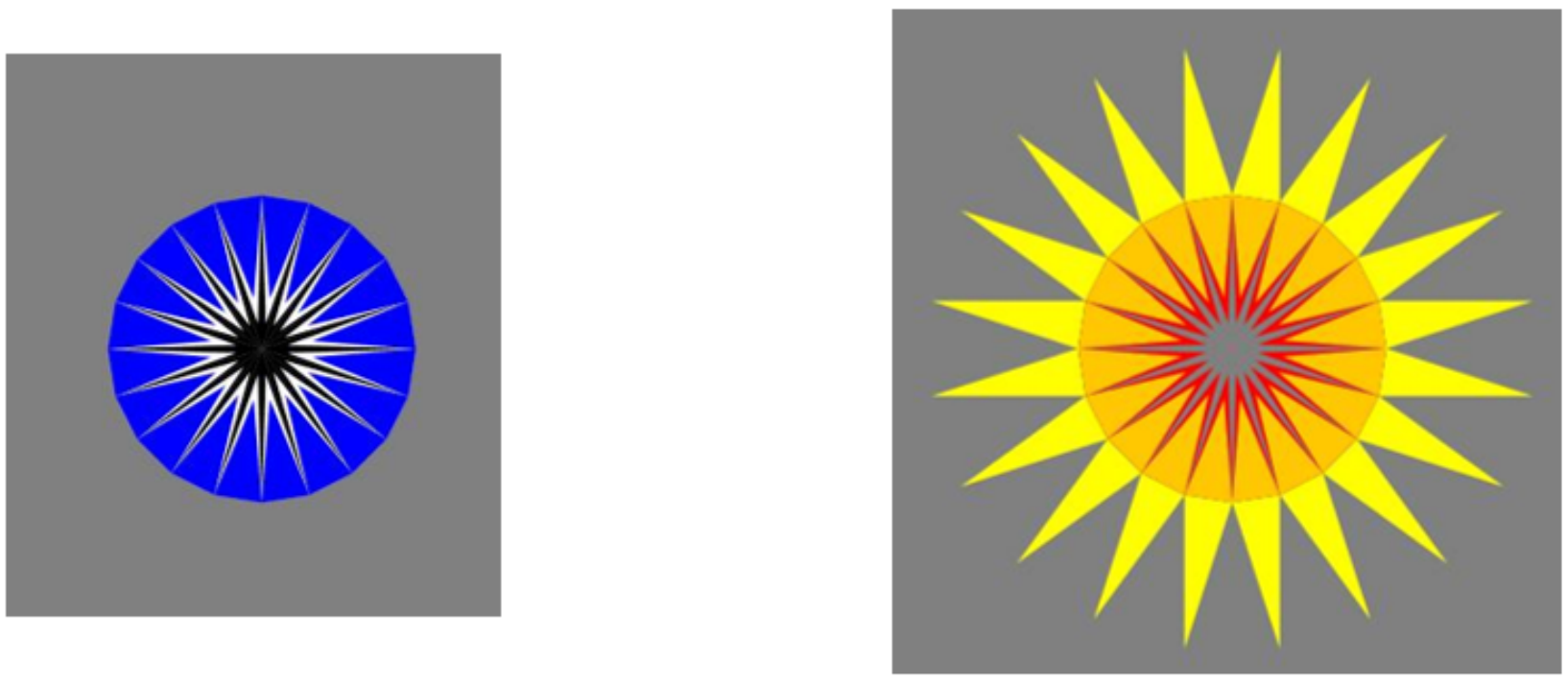

**Figure 10** Drawing stars (left picture) and suns (right picture) through a single transformation and color changes

Another example, illustrated in Figure 11, involves two versions of the Olympic rings, constructed using methods, again from *StdDraw*, as well as from an in-house library, *CircleDefs*, that includes methods for drawing circle based pictures; the *CircleDefs* methods are defined using the basic methods of *StdDraw* for plotting points or drawing straight line segments. In the Olympic rings exercise, the given pictures can be accompanied with the playing of the Olympic anthem. Putting such background sound is made easy through the use of the *StdAudio* library of the course textbook. Students are also asked to compare and contrast discrete and graphical creations of the UCY logo; Figure 12 illustrates such pictures. Regarding the discrete picture, what is shown here is a simple version. Often, other much more intricate creations are designed and drawn by students.The graphical drawings of the logo sometimes include animation as well.

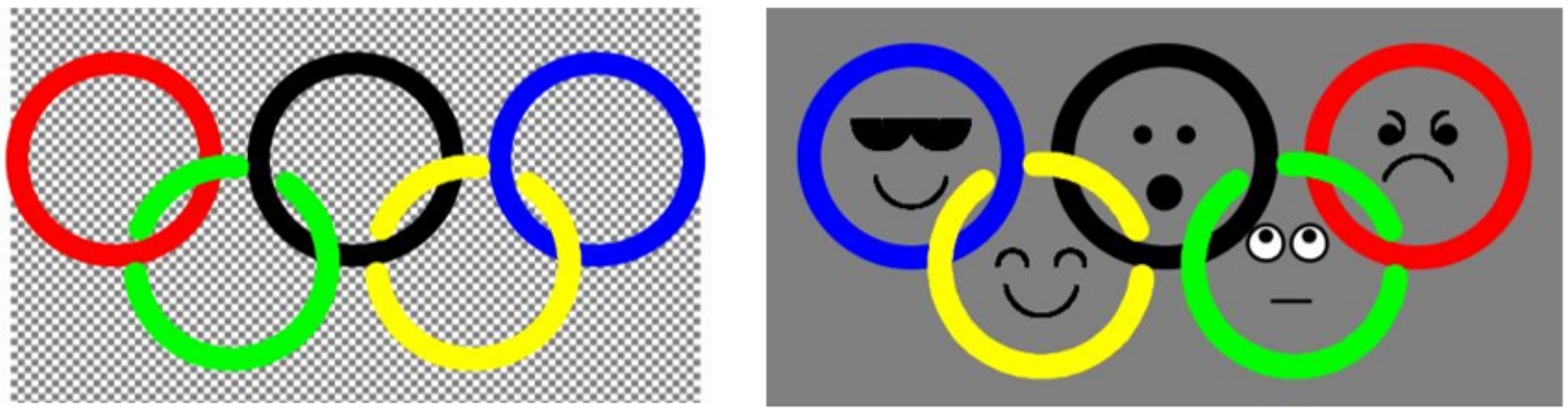

**Figure 11** Two versions of the Olympic rings

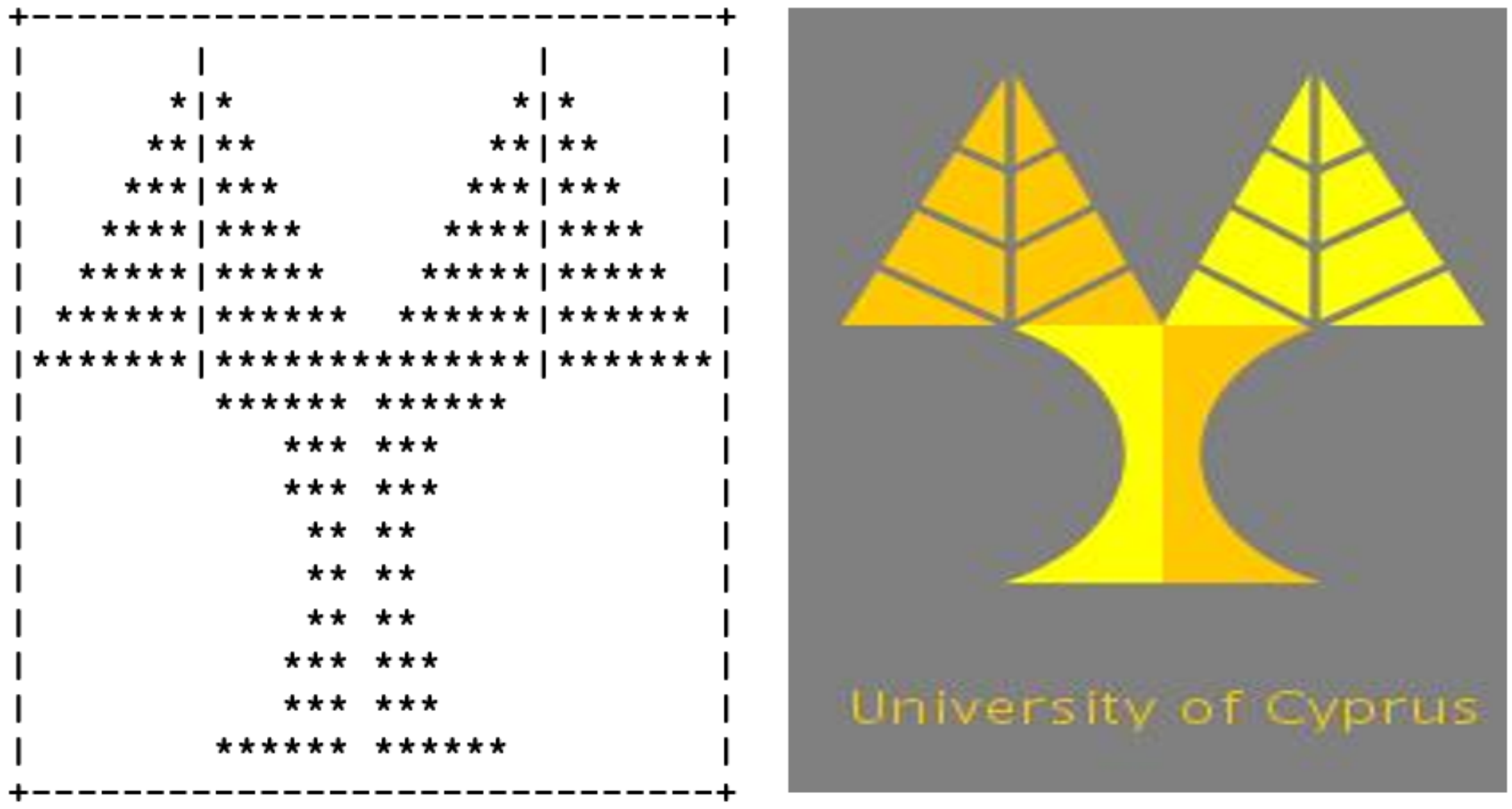

**Figure 12** A discrete and graphical version of the UCY logo

With respect to other graphical pictures, students can define moving turtles, bouncing balls, charges and their potentials, robots, etc., visualize, animate them or use them to draw other pictures in the case of turtles. The visual domain is a bottomless source for new drawing inspirations and programming exercises.

## 4 Computability Issues and the Visual Domain

The visual domain can also be utilized in airing computability issues with the students. We consider this an important added value of the presented visual approach. The students, by observing the construction of graphical drawings, can get easily acquainted in a visual way, with computation complexity issues. One such computation complexity exercise concerns the drawing of a filled circle, in at least two different ways, e.g. concentrically and radially (the given methods are included in the *CircleDefs* library) and observing the difference in the respective computation times which is substantial. Figure 13(a) shows a radially constructed filled circle on the left column, and three concentrically constructed versions on the right column. The bottom picture of these three is the one that is comparable to the radial one. In this example the primitive construct is plot a line. The radial picture on the left needed the plotting of just 180 lines, while the concentric pictures on the right from top to bottom respectively needed the plotting of 1800, 7200 and 28800 lines. The difference between the 180 and the 28800 lines is clearly evident, simply by observing the drawing of the respective pictures. The students are also asked to think of other ways of constructing filled circles (using lines or points) and to measure and plot, as shown in Figure 13(b), the computation time (y-axis) for different circle radius (x-axis). The graphical plotting of the computation time per se is also a visual drawing exercise. The higher interpretability of the graphical plot is compared against the tabular representation of the given times and radii in terms of numbers.

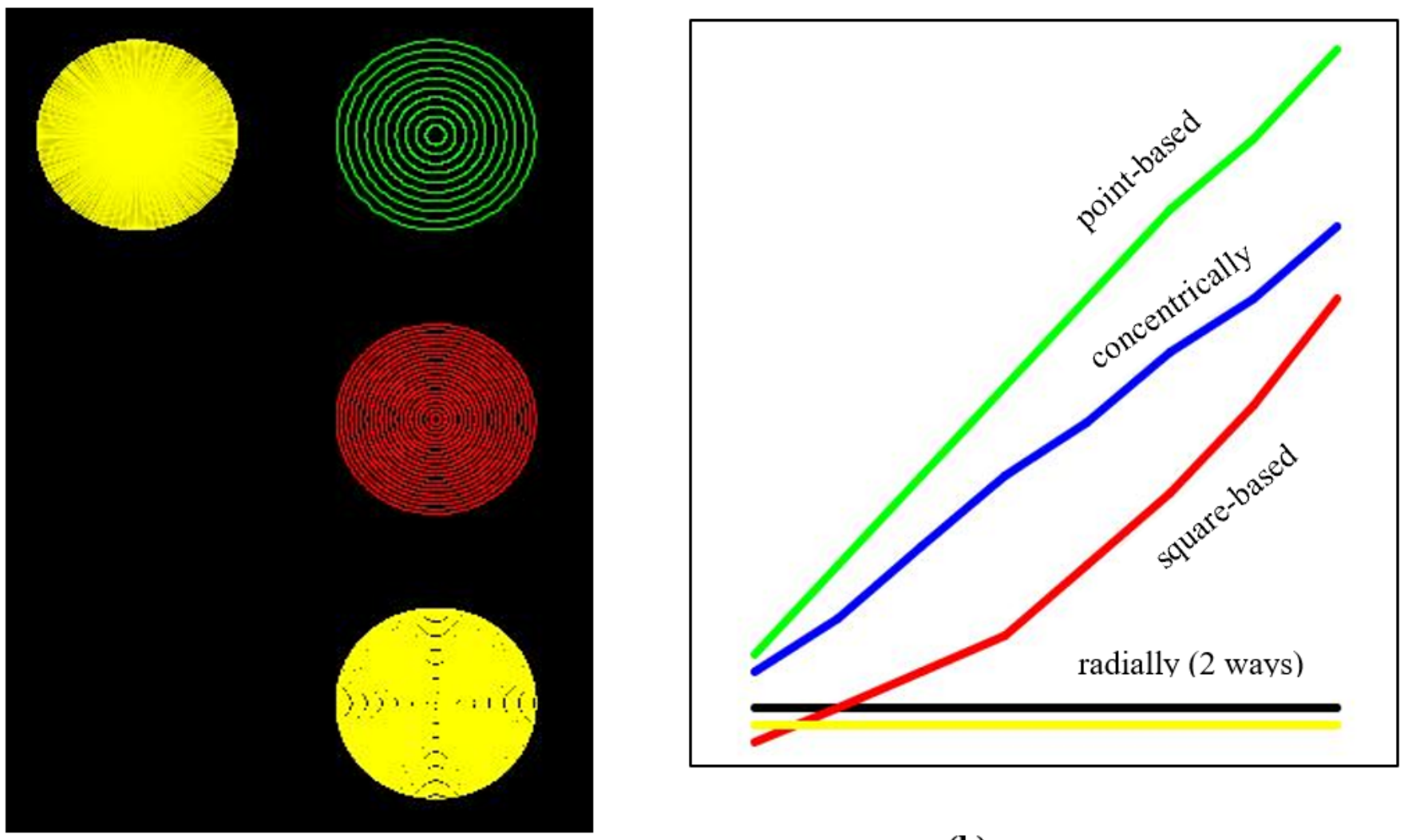

(b)

(a)

**Figure 13** (a) Drawing a filled circle, concentrically and radially and comparing in real-time the computation time; (b) Measuring and plotting computation times for different circle drawing methods and circle sizes

Measuring the complexity of a number of sorting algorithms empirically and plotting the graphs resulting from these measurements is another programming exercise. Moreover, complexity issues are extensively discussed with respect to recursion. For example, the complexity of stack recursive, tail recursive and iterative methods for computing the $n^{th}$ term of the Fibonacci series, is discussed and the relevant graphs for computing each of the first 30 terms of the series are produced. This goes to show the complexity equivalence between the tail recursive and iterative methods, and to further argue that complexity issues should not necessarily be advocated against the use of recursion if a tail recursive definition can replace a computationally (prohibitively) heavy stack recursive variant of some method, especially if recursion is the natural way to express the particular computation. This invariably triggers further discussion on the simplicity and comprehensibility of code that one should always strive to achieve.

## 5 Concluding Remarks and Further Endeavors

This paper is an essay about programming that aims to share with a broader audience of computer science educators and students our experience in utilizing the visual domain of pictures, in the context of our programming principles course taught to Computer Science majors in their first semester of studies. The particular course has substantially evolved through its thirty years of delivery, particularly regarding the underlying programming paradigm, which initially was the functional paradigm, then it was the procedural paradigm and now it is a synergistic interplay of the procedural and object-oriented paradigms, focusing on process and concept based thinking and modeling. However, a standard pedagogical feature of the course has been the utilization of the visual domain of discrete figures and graphical drawings, capitalizing on the structural aspects of all kinds of such pictures. Moreover, the domain of pictures is fairly neutral towards potential backgrounds of incoming students and as such it gives no competitive advantage to students with a specific background.

A current, related, strategic objective of CS-UCS is to enhance the soft skills of its graduates and to close the soft skills gap, through specific concerted actions in direct collaboration with industry [25], where an Industrial Advisory Board was recently formed. Coupled with this objective, a walk-in lab has been developed and added to the learning resources of CS-UCY from the academic year 2019/20. Before the pandemic the walk-in lab had been used for some of the tutorial sessions of the programming principles course for group-based problem solving, and hopefully it will be possible to resume this practice soon. In addition, before the pandemic we experimented with pair programming [13, 41], where the pairs of students were formed on the basis of academic ability as demonstrated by the mid-semester examination results (this is why the pairing was delayed until after the mid semester examination). Thus pairs consisted of students of comparable academic ability. We did not experiment with mixed ability pairs (i.e. a high achiever and another student of much lower ability), as over the years we have observed that often when students form working pairs, purely on their own initiative, the members of such pairs are of a comparable academic ability. Very strong students are not usually involved in such working pairs, and indeed in our limited experience with pair programming, it was observed that such students participated in the experiment quite reluctantly and their exchanges with their pairs were kept to the minimum required; it appears that students very strong in computer coding tend to be sole workers. Overall, the student comments received on this experience were quite positive and thus it is planned to resume and enhance the use of pair programming, even remotely, in the laboratory sessions of the course. Given that a number of other active learning methods of relevance to computing education impinge directly on problem solving and programming [18, 20, 21, 29, 34, 43, 51], the next evolution of the programming principles course is foreseen to be more towards the direction of active learning on a par with the development of important soft skills, where students will still exert their artistic skills through figuring and drawing, but rather in a collaborative fashion. We'll continue, though, to have competitions on student creations, assessed by the students themselves as currently done. Last but not least our current efforts in increasing the percentage of female students and bringing it back to what it used to be about 10 years ago, when we had a fairly balanced situation, will continue and be further enhanced; in this context we plan to carry out our own experiments trying to corroborate, or not, findings reported in the literature about the interest of female students both in the use of the visual domain in programming classes and in collaborative problem solving. Hopefully these developments will give rise to new insights worth reporting on, in a future article.

**Acknowledgements** The author is greatly indebted to all colleagues who taught the CS131 course at some stage, for the many exchanges and the numerous insights shared, in particular Ch. Georgiou, Y. Chrysanthou, Y. Sazeidis and G. Pallis, and the special teaching staff P. Antoniou and B. Bratskas. Last but not least the contribution of the hundreds of students who through their course evaluations and other feedback, helped in improving the course, as well as the help of the graduate students supporting the course as teaching assistants, are acknowledged. The insightful comments, suggestions and overall critique of the reviewers resulted in substantially improving the content of the paper.

## References

[1] Stephen Adam. "An introduction to learning outcomes: a consideration of the nature, function and position of learning outcomes in the creation of the European Higher Education Area". In: *EUA Bologna Handbook* B2.3-1 (2006).

[2] Stephen Adam. "Learning outcomes current developments in Europe: Update on the issues and applications of learning outcomes associated with the Bologna Process". In: *Bologna Seminar: Learning outcomes based higher education: the Scottish experience*. Volume 21. 22.02. 2008.

[3] Mordechai Ben-Ari. "Constructivism in computer science education". In: *ACM SIGCSE Bulletin* 30.1 (1998), pages 257–261. DOI: 10.1145/274790.274308.

[4] Jens Bennedsen and Michael E. Caspersen. "Programming in context: a model-first approach to CS1". In: *Proceedings of the 35th SIGCSE technical symposium on Computer science education*. 2004, pages 477–481. DOI: 10.1145/971300.971461.

[5] Richard Bird and Philip Wadler. *Introduction to Functional Programming. Series in Computer Science*. Prentice-Hall International, 1988. ISBN: 978-0134841892.

[6] Richard Bornat. *Programming from first principles*. Prentice Hall International (UK) Ltd., 1987. ISBN: 978-0137291045.

[7] Peter Brusilovsky, Eduardo Calabrese, Jozef Hvorecky, Anatoly Kouchnirenko, and Philip Miller. "Mini-languages: a way to learn programming principles". In: *Education and information technologies* 2.1 (1997), pages 65–83.

[8] Computer Science Curricula. *Curriculum Guidelines for Undergraduate Degree Programs in Computer Science*. https://www.acm.org/binaries/content/assets/education/cs2013_web_final.pdf. 2013. [visited on 8-May-2021].

[9] James H. Davenport, Alan Hayes, Rachid Hourizi, and Tom Crick. "Innovative pedagogical practices in the craft of computing". In: *2016 International Conference on Learning and Teaching in Computing and Engineering (LaTICE)*. IEEE. 2016, pages 115–119.

[10] Todd Davey, Arno Meerman, Balzhan Orazbayeva, Max Riedel, Victoria Galán-Muros, Carolin Plewa, and Natascha Eckert. *The Future of Universities Thoughtbook: 40 Perspectives on how engaged and entrepreneurial universities will drive growth and shape our knowledge-driven future until 2040*. University Industry Innovation Network, 2018. ISBN: 978-94-91901-32-4.

[11] Anna Eckerdal, Michael Thuné, and Anders Berglund. "What does it take to learn'programming thinking'?" In: *Proceedings of the first international workshop on Computing education research*. 2005, pages 135–142.

[12] *ECTS Users' Guide*. https://ec.europa.eu/education/ects/users-guide/docs/ects-users-guide_en.pdf. 2015. [visited on 8-May-2021].

[13] Brian Hanks, Sue Fitzgerald, Renée McCauley, Laurie Murphy, and Carol Zander. "Pair programming in education: A literature review". In: *Computer Science Education* 21.2 (2011), pages 135–173. DOI: 10.1080/08993408.2011.579808.

[14] Jeri R. Hanly and Elliot B. Koffman. *Problem Solving & Program Design in C*. Addison Wesley Longman, Inc, 2010. ISBN: 9780321535429.

[15] Rachel Harrison. "The use of functional languages in teaching computer science". In: *Journal of Functional programming* 3.1 (1993), pages 67–75.

[16] Orit Hazzan. "How students attempt to reduce abstraction in the learning of mathematics and in the learning of computer science". In: *Computer Science Education* 13.2 (2003), pages 95–122.

[17] Rosario Hernández. "Does continuous assessment in higher education support student learning?" In: *Higher education* 64.4 (2012), pages 489–502. DOI: 10.1007/s10734-012-9506-7.

[18] Diane Horton, Michelle Craig, Jennifer Campbell, Paul Gries, and Daniel Zingaro. "Comparing outcomes in inverted and traditional CS1". In: *Proceedings of the 2014 conference on Innovation & technology in computer science education*. 2014, pages 261–266. DOI: 10.1145/2591708.2591752.

[19] Emily Howe, Matthew Thornton, and Bruce W. Weide. "Components-first approaches to CS1/CS2: principles and practice". In: *ACM SIGCSE Bulletin* 36.1 (2004), pages 291–295. DOI: 10.1145/971300.971404.

[20] Christopher D. Hundhausen, Anukrati Agrawal, and Pawan Agarwal. "Talking about Code: Integrating Pedagogical Code Reviews into Early Computing Courses". In: *ACM Transactions on Computing Education (TOCE)* 13.3 (2013). DOI: 10.1145/2499947.2499951.

[21] Ville Isomöttönen and Ville Tirronen. "Teaching programming by emphasizing self-direction: How did students react to the active role required of them?" In: *ACM Transactions on Computing Education (TOCE)* 13.2 (2013), pages 1–21. DOI: 10.1145/2483710.2483711.

[22] Elaine Kao. *Exploring Computational Thinking. Google AI Block*. https://ai.googleblog.com/2010/10/exploring-computational-thinking.html. 2010. [visited on 8-May-2021].

[23] Declan Kennedy. *Writing and using learning outcomes: a practical guide*. University College Cork, 2006. ISBN: 9780955222962.

[24] Elpida T. Keravnou. "Introducing computer science undergraduates to principles of programming through a functional language". In: *International Symposium on Functional Programming Languages in Education*. Springer. 1995, pages 15–34.

[25] Elpida Keravnou-Papailiou. "Attacking the soft skills gap by placing industrial internships on a new footing in our Computer Science curriculum". In: *University Industry Interaction Conference*. 2019.

[26] Elpida Keravnou-Papailiou. "Flexibility through Learning Outcomes: Implications for Quality". In: *Proc. Third European Quality Assurance Forum*. 2008.

[27] Elpida Keravnou-Papailiou. "Implementing ECTS at the University of Cyprus". In: *EUA Bologna Handbook* C3.3-2 (2007), pages 1–22.

[28] Jeff Kramer. "Is abstraction the key to computing?" In: *Communications of the ACM* 50.4 (2007), pages 36–42.

[29] Clifton Kussmaul. "Process oriented guided inquiry learning (POGIL) for computer science". In: *Proceedings of the 43rd ACM technical symposium on Computer Science Education*. 2012, pages 373–378.

[30] Greg Michaelson. "Microworlds, objects first, computational thinking and programming". In: *Computational thinking in the STEM disciplines: Foundations and Research Highlights*. Springer, 2018, pages 31–48. ISBN: 9783319935669.

[31] Greg Michaelson. "Programming Paradigms, Turing Completeness and Computational Thinking". In: *The Art, Science, and Engineering of Programming* 4.3 (Feb. 2020), 4:1–4:21. ISSN: 2473-7321. DOI: 10.22152/programming-journal.org/2020/4/4.

[32] Greg Michaelson. "Teaching programming with computational and informational thinking". In: *Journal of Pedagogic Development* (2015).

[33] James Millen, Louise McDowell, and Pamela Larmour. *Why Don't More Young Women Study Computing? A working paper investigating the low participation of girls taking computing in Northern Ireland*. Council for the Curriculum, Examinations and Assessment (CCEA) Research & Statistics Unit, 2019.

[34] Rakesh Mohanty and Sudhansu Bala Das. "A proposed what-why-how (WWH) learning model for students and strengthening learning skills through computational thinking". In: *Progress in Intelligent Computing Techniques: Theory, Practice, and Applications*. Springer, 2018, pages 135–141. DOI: 10.1007/978-981-10-3376-6_15.

[35] Ellen Murphy, Tom Crick, and James H Davenport. "An Analysis of Introductory Programming Courses at UK Universities." In: *The Programming Journal* 1.2 (2017), page 18. DOI: 10.22152/programming-journal.org/2017/1/18/.

[36] Claudia Ott, Anthony Robins, and Kerry Shephard. "Translating principles of effective feedback for students into the CS1 context". In: *ACM Transactions on Computing Education (TOCE)* 16.1 (2016), pages 1–27.

[37] Tauno Palts and Margus Pedaste. "A model for developing computational thinking skills". In: *Informatics in Education* 19.1 (2020), pages 113–128.

[38] Arnold Pears, Stephen Seidman, Lauri Malmi, Linda Mannila, Elizabeth Adams, Jens Bennedsen, Marie Devlin, and James Peterson. "A Survey of Literature on the Teaching on Introductory Programming". In: *ACM SIGCSE Bulletin* 39.4 (2007), pages 204–223. DOI: 10.1145/1345375.1345441.

[39] Keith Quille, Susan Bergin, and Aidan Mooney. "Programming: Factors that influence success revisited and expanded". In: *International Conference on Engaging Pedagogy (ICEP), 3rd and 4th December, College of Computing Technology, Dublin, Ireland*. Volume 10. 2015, pages 1047344–1047480.

[40] Atanas Radenski. "Digital support for abductive learning in introductory computing courses". In: *Proceedings of the 38th SIGCSE technical symposium on Computer science education*. 2007, pages 14–18. DOI: 10:1145/1227310.1227318.

[41] Alex Radermacher, Gursimran Walia, and Richard Rummelt. "Improving student learning outcomes with pair programming". In: *Proceedings of the ninth annual international conference on International computing education research*. 2012, pages 87–92. DOI: 10.1145/2361276.2361294.

[42] Lauren Rich, Heather Perry, and Mark Guzdial. "A CS1 course designed to address interests of women". In: *Acm sigcse bulletin* 36.1 (2004), pages 190–194. DOI: 10.1145/971300.971370.

[43] Kate Sanders, Jonas Boustedt, Anna Eckerdal, Robert McCartney, and Carol Zander. "Folk pedagogy: Nobody doesn't like active learning". In: *Proceedings of the 2017 ACM Conference on International Computing Education Research*. 2017, pages 145–154.

[44] Robert Sedgewick and Kevin Wayne. *Introduction to programming in Java: an interdisciplinary approach*. Addison-Wesley, 2017. ISBN: 9780321498052.

[45] Beth Simon, Päivi Kinnunen, Leo Porter, and Dov Zazkis. "Experience report: CS1 for majors with media computation". In: *Proceedings of the fifteenth annual conference on Innovation and technology in computer science education*. 2010, pages 214–218.

[46] Codementor Team. *What programming languages should a beginner learn in 2019?* https://www.codementor.io/codementorteam/beginner-programming-language-job-salary-community-7s26wmbm6. 2019. [visited on 8-May-2021].

[47] Arto Vihavainen, Jonne Airaksinen, and Christopher Watson. "A systematic review of approaches for teaching introductory programming and their influence on success". In: *Proceedings of the tenth annual conference on International computing education research*. 2014, pages 19–26.

[48] Daiva Vitkutė-Adžgauskienė and Antanas Vidziunas. "Problems in choosing tools and methods for teaching programming". In: *Informatics in education* 11 (2012), pages 271–282.

[49] Brenda Cantwell Wilson and Sharon Shrock. "Contributing to success in an introductory computer science course: a study of twelve factors". In: *ACM SIGCSE Bulletin* 33.1 (2001), pages 184–188. DOI: 10.1145/366413.364581.

[50] Jeannette M. Wing. "Computational thinking and thinking about computing". In: *Philosophical Transactions of the Royal Society A: Mathematical, Physical and Engineering Sciences* 366.1881 (2008), pages 3717–3725.

[51] Krissi Wood, Dale Parsons, Joy Gasson, and Patricia Haden. "It's never too early: pair programming in CS1". In: *Proceedings of the Fifteenth Australasian Computing Education Conference-Volume 136*. 2013, pages 13–21.

## About the author

**Elpida Keravnou-Papailiou** is Professor of Computer Science at the University of Cyprus and Departmental Chair. Previously she served her university from various academic administrative positions (Dean of the Faculty of Pure and Applied Sciences and Vice-Rector of Academic Affairs during which time she coordinated the implementation of the Bologna reforms at the University of Cyprus). Although her research is in Artificial Intelligence, her interest in programming started from the beginning of her academic career at the Department of Computer Science of University College London where she taught a course on Special Purpose Programming Languages. This interest was strengthened substantially when she undertook the teaching of programming principles to new coming students at the Department of Computer Science of the University of Cyprus. Contact her at elpida@ucy.ac.cy.